\documentclass{aa}  
\usepackage[colorlinks, citecolor=blue]{hyperref}
\usepackage{soul}
\usepackage{natbib}
\bibpunct{(}{)}{;}{a}{}{,} 

\usepackage{graphicx}
\usepackage{float}
\usepackage{caption}
\usepackage{subcaption}
\usepackage{afterpage}
\usepackage{txfonts}

\begin{document}

   \title{Vertical and radial distribution of atomic carbon in HD 163296}

   \author{F. Urbina
          \inst{1}
          \and
          J. Miley \inst{2,3,4}
          \and M. Kama 
          \inst{5,6}
          \and L. Keyte
          \inst{5}}

   \institute{University of Chile, Astronomy Deparment, Camino El Observatorio 1515, Santiago, Chile\\
              \email{francisco.urbina@ug.uchile.cl}
         \and
             Departamento de Física, Universidad de Santiago de Chile, Av. Victor Jara 3659, Santiago, Chile
        \and Millennium Nucleus on Young Exoplanets and their Moons (YEMS), Chile
        \and Center for Interdisciplinary Research in Astrophysics and Space Exploration (CIRAS), Universidad de Santiago, Chile
        \and Department of Physics and Astronomy, University College London, Gower Street, London, WC1E 6BT, UK
        \and Tartu Observatory, University of Tartu, Observatooriumi 1, T\~{o}ravere, 61602, Estonia\\
             }

   \date{Received November 9, 2023; accepted March 1, 2024}

 
  \abstract
   {In protoplanetary disks, atomic carbon is expected to originate from the PDR at the disk surface where CO is dissociated by UV photons coming from the stellar, or external interstellar, radiation field. Even though atomic carbon has been detected in several protoplanetary disks, there is a lack of spatially resolved observations of it. }
  {For HD 163296 protoplanetary disk, we aim to obtain both  radial and vertical structure of [CI]$={^3}{P_1} - {^3}{P_0}$ line emission and perform the first direct comparison of this tracer with optically thick line emission $^{12}$CO $J=2-1$.}
   {We used archival ALMA data for [CI]$={^3}{P_1} - {^3}{P_0}$ and previously published ${^{12}}$ CO $J=2-1$ data in HD 163296. Through the software of disksurf we extracted the vertical structure, meanwhile radial profiles were obtained directly from imaging. Brand new DALI modelling was employed to perform direct comparison to the data.}
   {We found that these tracers are collocated radially but not vertically, where $^{12}$CO $J=2-1$ emission is, on average, located at higher altitudes, as it is also the case for other tracers in the same disk. }
   {Due to this difference in vertical height of the emission, the optically thick $^{12}$CO $J=2-1$ emission line appears to trace the highest altitudes, despite the expected formation mechanism of [CI] in the disk. The latter phenomena may be due to efficient mixing of the upper layers of the disk, or UV photons penetrating deeper than we expected. }

   \maketitle
%

\section{Introduction}

Our knowledge about protoplanetary disks compositions and structure has been greatly expanded by  high resolution observations of the Atacama Large Millimeter and sub-millimeter Array (ALMA). For example, the DSHARP project \citep{DSHARP},   unveiled disks' dust substructures using high resolution observations. Meanwhile the MAPS large program investigated the gas in protoplanetary disks in order to study their chemical structure \citep{MAPS}. As well as investigating the radial structure of disks, we now have the instrumental capability to enable studies of the vertical structure, for which different approaches have been taken.
For dust, observing nearly edge-on systems \citep{villenave2020observations} or analyzing scattered light observations (see  \cite{avenhaus2018disks} for an example) are both methods for assessing the vertical structure of disks. Meanwhile, for the gas, given sufficient spectral and spatial resolution, it is possible to infer the vertical structure from the displacement of the emission relative to the perfectly flat disk case \citep{2021A&A...650A.179I}. One way is to fit velocity curves, as done with ConeRot \citep{casassus2019kinematic}, which places the emission in a conical surface and makes use of it to define average velocity curves which are then use to fit the geometry of such conical surface, enabling vertical structure detection. A method that is non-model dependent and well suited for mid-inclination disks was first presented in \cite{pintemethod}. In a similar fashion as in ConeRot,  this method uses a geometrical approach by placing the channel maps emission in a conical surface ,where the position of the emission is estimated by the local maximum of it, and has been successfully applied to constrain, for example, carbon monoxide vertical structure in several systems \citep{law2022co}.

Atomic carbon, [CI], is a useful tracer for studying the carbon content in protoplanetary disks. Typical models for [CI] in circumstellar disks consider it as a product of the photo dissociation of carbon monoxide by UV photons, therefore, it is expected for atomic carbon to trace the upper layers of the disk, where the ionizing radiation is abundant \citep{kama2016observations}. As a consequence, [CI] should act as a reliable tracer for the vertical height 
 in protoplanetary disks. Some observations towards debris disks suggest [CI] has an important shielding effect on CO protecting it from UV photons and forming a layer close to it in the photo dissociation region (PDR), sharing a similar phenomenology as the interstellar medium \citep{van1988photodissociation}. In spatially resolved observations of debris disks, [CI] appears to be co-located radially though not vertically \citep{higuchi2019first, kral2019imaging}.

\begin{table*}[t]
		\centering
		\caption{Summary of the noise RMS levels, peak SNR and 3$\sigma$ integrated flux from [CI] channels maps, [CI] zeroth moment map.}
		\begin{tabular}{cccc}
			\hline
			\hline
			&\textbf{[CI] channel maps} &  \textbf{[CI] zeroth moment map}  & \textbf{Continuum}  \\
			\hline
		
            Noise RMS (Jy Beam$^{-1}$) &   2.16$\times 10^{-2}$ &  2.63 $\times 10^{-2}$ (km/s)      & $6.62 \times 10^{-4}$  \\
		Peak SNR  & 14.26       &      17.81 &   1324.93 \\
		Integrated flux  & - & 7.24 $\pm$ $0.02$ (Jy km/s) & 3.51 $\pm$   $0.04 \times 10^{-2}$ (Jy)\\

			\hline
		\end{tabular}
		\label{tab:table measurements}
	\end{table*}

[CI] has been detected towards protoplanetary disks in emission using single dish observations by APEX \citep{kama2016observations}. There, 6 out 12 disks showed [CI]$={^3}{P_1} - {^3}{P_0}$ line emission detection, meanwhile the [CI]$={^3}P_2 - {^3}P_1$ line turned out to be fainter as only 1 system out the 33 surveyed were detected at that frequency. For systems like TW Hya and HD 100546 it was suggested that carbon is under-abundant \citep{bruderer2012warm, du2015volatile, kama2016volatile}, due to processes like freeze-out of molecules in the mid plane of the formation of complex organic molecules in cold regions. Nevertheless these unresolved observations were not able to precisely constrain the gas-phase [C]/[H] ratio due to lack of sensitivity. To test the vertical structure, spatially resolved observations for [CI]$={^3}{P_1} - {^3}{P_0}$ are needed but there is still a lack of them. To the authors knowledge, there are only two published examples of spatially resolved imaging of [CI] towards a Class II protoplanetary disks. In \citep{Imlup_recent}, IM Lup [CI] emission was examined,  finding this emission is tracing the upper layers of the disk, even at higher altitudes relative to CO. Meanwhile in \cite{https://doi.org/10.48550/arxiv.2211.16531} a kinematic bipolar structure was detected in the [CI] emission of HD 163296, possibly associated with a planet candidate, but no further analysis to the radial and vertical structure of the emission was done.

  HD 163296 protoplanetary disk has been exhaustively studied, it has a rich dust structure showing gaps, rings and a crescent \citep{isella2018disk}. Its vertical structure has been studied for several tracers such as CO and its isotopologues and other molecules \citep{Tere}. To date, the most prominent  vertical structure found is detected for $^{12}$CO $J=2-1$ line emission, even when compared to its isotopologues and other different tracers, which is explained by the fact emission lines from molecules such as HCN, CN or isotopologues of CO are optically thin.

In this work, we aim to test our current knowledge about the expected location of atomic carbon within protoplanetary disks by directly comparing the radial and vertical location of [CI]   and CO in HD 163296. The paper is organized as follows. In Section \ref{sec:observations}, we describe the procedure for reduction and imaging of the archival ALMA data used in this work. The analysis tools and results are presented in Section \ref{sec:results}. In Section \ref{sec:disucussions} we compare the observed structure as indicated by [CI] and CO emission to modelling.  Finally, section \ref{sec:summary} summarizes the important analyzes and discussions from this work.   

\section{Observations and data reduction}
\label{sec:observations}

    We used archival ALMA band 8 observations of HD163296 from the project 2013.1.00527.S (PI: M. Kama). These observations  were obtained in a single pointing on May 2, 2015, affected by an average precipitable water vapor of 0.536 mm. A configuration of 37 antennas of 12 m where the projected baselines 5th and 80th percentile are 9.92 m and  127.36 m respectively was used, which resulted in a synthesizing beam for the continuum images of 0.675$\times$0.296" when constructing images with natural weighting. This weighting was chosen to compensate for the low sensitivity of the observations as the emission is expected to be faint and the total integration time on source was only 13.1 minutes.  Dust continuum was observed in 2 spectral windows, each one with a bandwidth of 2 GHz, 128 channels and central frequencies of 480.24 and 482.05 GHz respectively. [CI] $={^3}P_1 - {^3}P_0$ line emission was observed in a single spectral window of 3840 channels of spacing 488.28 KHz, with a central frequency of 492.22 GHz. Calibration sources were the quasars J1256-0547 (bandpass) and J1733-1304 (phase) and Titan (flux).
\begin{figure}[h]
    \centering
    \includegraphics[scale=0.55]{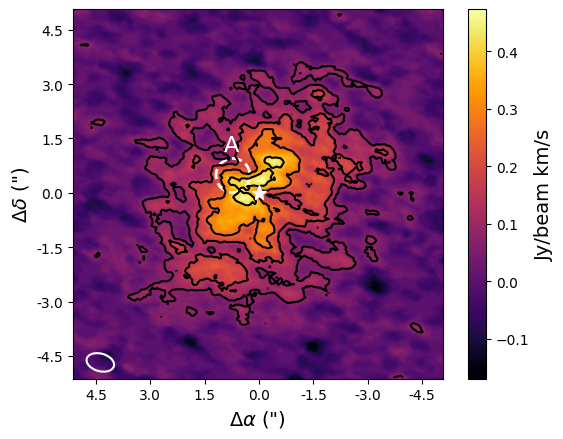}
    \caption{HD163296 ALMA band 8 zeroth moment map of [CI] line emission. Bottom left white ellipse shows the synthesized beam FWHM. Star position is indicated at $(0,0)$ by a white figure. Contours show 3$\sigma$, 6$\sigma$, 10$\sigma$ and 15$\sigma$ (1$\sigma=2.63 \times 10^{-2}$ Jy/beam km/s). Only the emission within velocity ranges from 1.3 to 10.6 km/s and pixels with values higher than $2\sigma$ where used. Enclosed region, labelled by letter A, highlights the position of an asymmetry feature detected in the channels maps (see Fig. \ref{fig:channels}). }
    \label{fig:zeroth}
\end{figure}

\begin{figure}[h]
    \centering
    \includegraphics[scale=0.55]{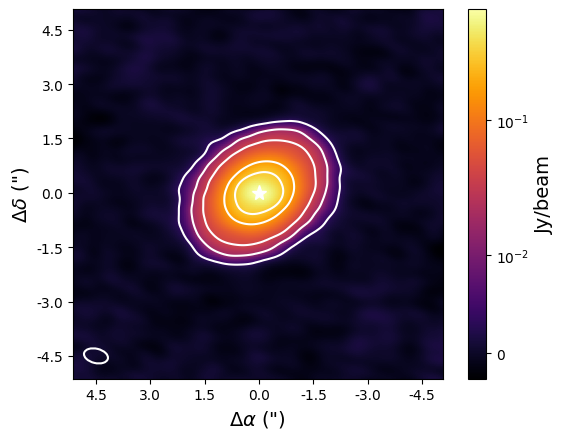}
    \caption{HD163296 ALMA band 8 dust continuum emission. Bottom left white ellipse shows the synthesized beam FWHM. Star position is indicated at $(0,0)$ by a white figure. Contours show 5$\sigma$, 20$\sigma$, 50$\sigma$ and 250$\sigma$ (1$\sigma = 6.62 \times 10^{-4}$ Jy/beam).}
    \label{fig:continuum}
\end{figure}

\begin{figure*}[h]
\centering
\begin{subfigure}{.5\textwidth}
  \centering
  \includegraphics[width=0.9\linewidth]{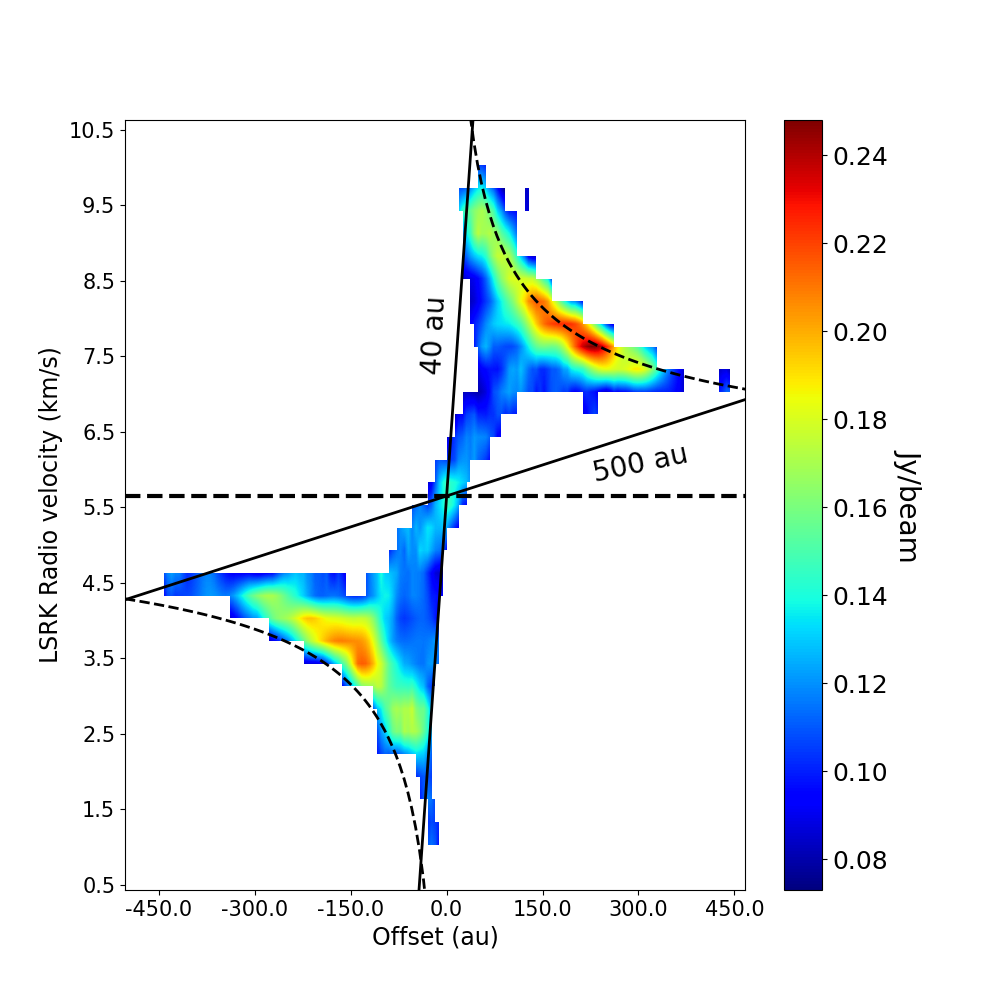}
\end{subfigure}%
\begin{subfigure}{.5\textwidth}
  \centering
  \includegraphics[width=.9\linewidth]{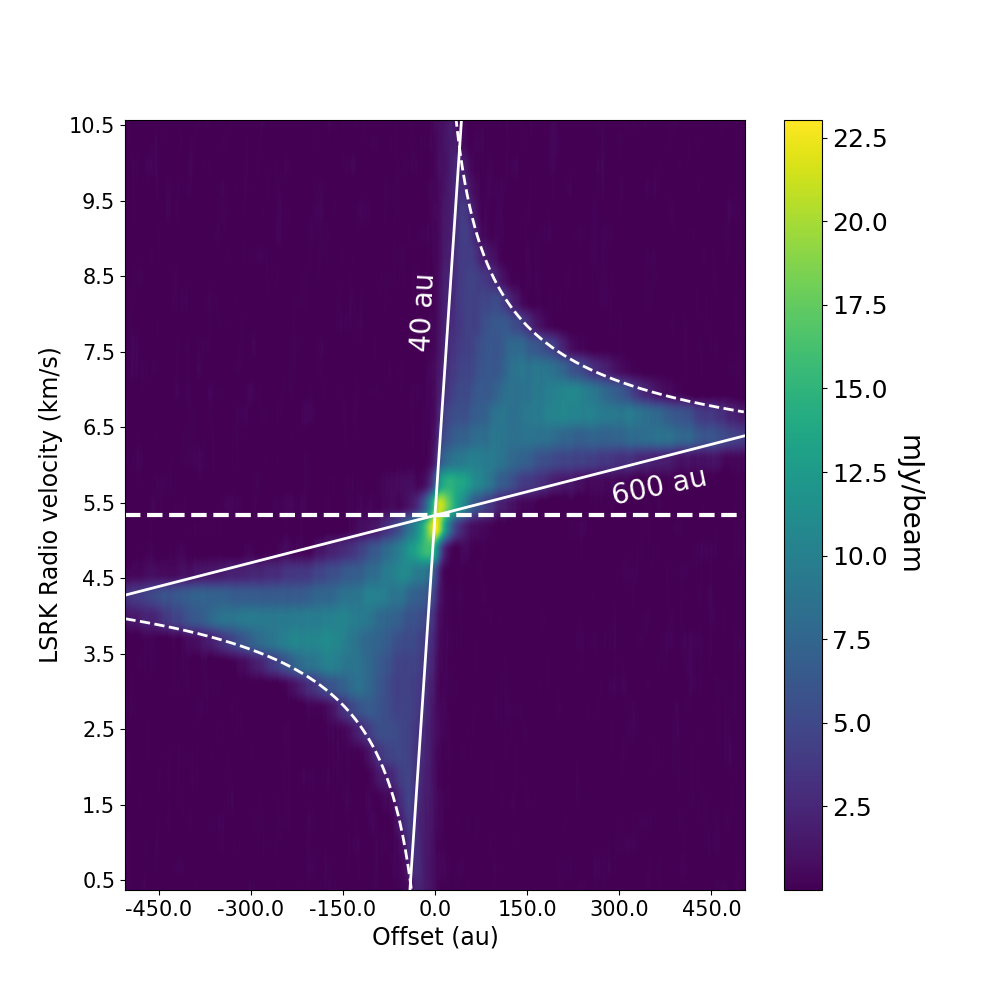}
\end{subfigure}
\caption{PV diagrams for CO and [CI] emission in HD 163296. Diagram done using CASA task \textit{impv}. The position axis is the apparent major axis of the gas disk. The starting position is $ (\alpha_{J2000}, \delta_{J2000})=(17^h56^m21^s.548,-21^{\circ} 57'25''.902)$  while the end position is$ (\alpha_{J2000}, \delta_{J2000})=(17^h56^m21^s.001,-21^{\circ}57'19''.436)$. Only pixels with values higher than $3\sigma$ were used. Dashed horizontal line shows the system velocity, curved dashed line show maximum radial velocity according to eq. (\ref{max vel}) and straight black solid lines are obtained according to eq. (\ref{radial vel}) enclosing where the emission is.}
\label{fig: PV diagrams}
\end{figure*}
All data sets were calibrated and flagged using \textit{Common Astronomy Software Applications} (CASA) \citep{2007ASPC..376..127M} using the pipeline script delivered with the data. Self calibration was carried out reaching a final solution interval of 30s for phase calibration. Amplitude self calibration was also applied, with the solutions applied to line imaging which increased the final SNR by 20\% compared to non-selfcalibrated products. Line imaging of [CI] was achieved after subtracting the continuum and averaging the spectral windows to a width of 0.3~km/s. The synthesised beam of the CI cube is $0.78" \times 0.49"$ with a PA of 74.46°. 

The data we present here is complemented by observations of $^{12}$CO $J=2-1$ towards HD~163296, which where directly obtained from the publically available DSHARP data release \citep{DSHARP,isella2018disk}.


\section{Analysis and results}
\label{sec:results}
 \subsection{Imaging}

In Fig. \ref{fig:channels} atomic carbon line emission channel maps are shown following the typical wing shape for gas rotating in a Keplerian fashion.   Fig \ref{fig:zeroth} presents the results of using the software \textbf{bettermoments} \citep{bettermoments} for obtaining the zeroth moment map of the emission by integrating from 1.3 to 10.6 km/s and masking all pixels with values lower than $2\sigma$.  In the moment 0 map more emission is observed towards the NE side of the disk as highlighted by the $15\sigma $ contours in the image.  Fig. \ref{fig:continuum} shows dust continuum imaging achieved by combining both dedicated spectral windows, with an achieved synthesized beam FWHM dimensions of $0.67" \times 0.40"$ and a PA of 76.60°. Table \ref{tab:table measurements} summarizes the noise levels and integrated fluxes.

\begin{figure}[h]
    \centering
    \includegraphics[scale=0.28]{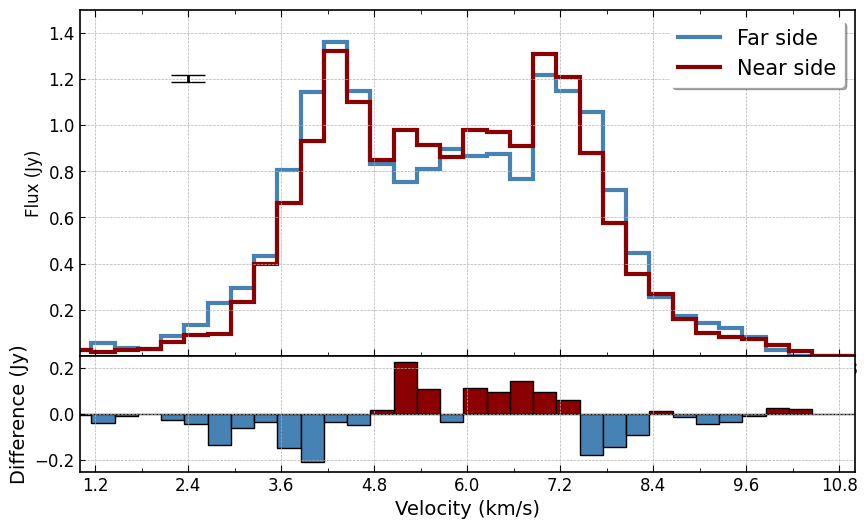}
    \caption{Extracted spectrum from the near and far side of the disk from the channel maps shown in Fig. \ref{fig:channels}. A flux excess coming from the near side of the disk is observed from 5.2 to 7.3 km/s. Top left black bar shows an estimate of the propagated error in both spectra. }
    \label{fig:assymetry}
\end{figure}

The asymmetry in the zeroth moment map is also shown in Fig. \ref{fig:channels}  where the northern side is brighter at channels around 5.5 km/s where we highlight region A showing an asymmetric peak. The NE of the disk is thought to be the near side of the disk as scattered light imaging indicates \citep{2019ApJ...875...38R}. To further investigate, spectra from the near and far side of the disk are extracted. This is achieved by dividing each channel along the major axis of the disk. Only emission above 3$\sigma$ is used and Keplerian masking is applied by using the \cite{rich_teague_2020_4321137} implementation. The extracted spectra can be seen in Fig. \ref{fig:assymetry} where an overall flux excess is observed from 5.2 to 7.3 km/s. The maximum excess comes from the channel at 5.5 km/s that can be due to the previously mentioned region A.

An apparent flux surplus coming from the far side of the disk is also measured towards the channels at velocities of 4.0 and 7.6 km/s. At those channels, the emissions wings are no longer spatially separated, and so distinguishing the near and far side in images becomes more difficult for the achieved angular resolution. Close to the major axis, the area of one synthesized beam can easily include emission from both the near and far side (and potentially the upper and lower side, see \cite{rosenfeld2013spatially}). We defer interpretation of the asymmetry in these specific channels until in depth modelling of CI emission processes applied to HD 163296 can be achieved. A well defined vertical height will be crucial for enabling these efforts.

\subsection{Radial analysis}

Figure \ref{fig: PV diagrams} show Position-Velocity (PV) diagrams for both CO and [CI] emissions in HD 163296 obtained along the major axis of the disk. For gas following Keplerian motion, a symmetric diagram around the origin is expected, as it is the case for the CO emission. The PV diagram of [CI] shows a minor asymmetry in the emission along the major axis, in which channels at velocities of  7.6-7.9 km/s are brighter than their symmetric counterparts at 3.7-4.0 km/s as can be seen in Fig. (\ref{fig:channels}), notice this is not the same asymmetry from the near-far side of the disk previously discussed and it is not related to the region A, as the latter does not belong to the major axis of the disk.

The inner and outer radii can be constrained by assuming Keplerian rotation of the gas (see \cite{matra2017exocometary} for an example).  The maximum radial velocity measured along the line of sight is

\begin{equation}
    v_{max} = \sqrt{\frac{GM_{*}}{x}} \sin(i)
    \label{max vel},
\end{equation}
where $x$ is the distance measured along the major axis of the disk, $i$ is the inclination of the disk, $M_*$ is the mass of the star and $G$ is the gravitational constant. Also, the radial velocity is 

\begin{equation}
    v_{rad} = \sqrt{\frac{GM_{*}}{R^3}} \sin(i)x,
    \label{radial vel}
\end{equation}
where $R$ is the radial distance from the star to the gas. With this method we find an inner radius of 40 au for both tracers and an outer radius of 600 au and 500 au for CO and [CI] respectively. The outer radius for CO is in good agreement with the value of 560 au reported in \cite{isella2018disk}. Notice the inner and outer radii for [CI] should be taken as an upper and lower limits for the actual values respectively, as these measurements are affected by the low sensitivity of the observations its effects cannot be ruled out with the current available data.

\begin{figure}[h]
    \centering
    \includegraphics[scale=0.37]{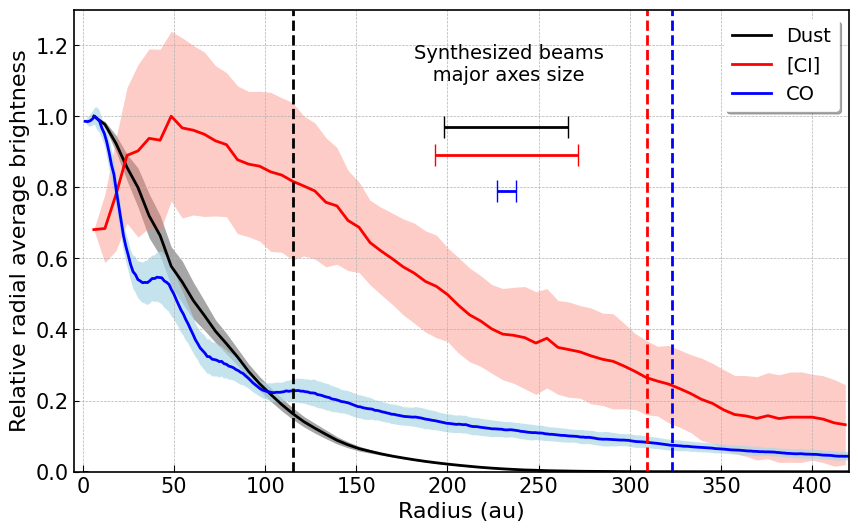}
    \caption{[CI], CO and dust continuum average radial profiles comparison for HD 163296. All profiles were normalized by its respective maximum value. Shaded shaded regions represent $\pm 1\sigma$ deviations from the average values. Color boxes show synthesized beam major axis for each dataset and dashed lines show $R_{90}$ \textit{cutoff}.  }
    \label{fig:radial avg}
\end{figure}

For comparing the radial extent of dust, atomic carbon and carbon monoxide, an azimuthal average profile is obtained for dust and line emissions moment 0 maps, as shown in Fig. \ref{fig:radial avg}.  A radial cutoff ($R_{90}$) is defined as the radius where 90\% of the total integrated profile value is reached. $R_{90}$ is measured for all three datasets studied. By sampling the radial profiles at each radius, several profiles are obtained for which the calculation is repeated. Their respective 68\% confidence interval is estimated. For [CI] and CO it is found $R_{90C} = 309.6^{+6.1}_{-12.1} $au and $R_{90CO} = 323.0^{+3.9}_{-2.7}$  au and for dust continuum $ R_{90 dust}= 115.14^{+6.1}_{-0.0}$ au. Therefore we obtain a ratio of $R_{90C}/R_{90dust} = 2.63^{+0.11}_{-0.13}$ and $R_{90CO}/R_{90dust} = 2.78^{+0.03}_{-0.05}$, suggesting dust drift, as expected for this system.

\begin{figure*}[h]
\centering
\begin{subfigure}{.5\textwidth}
  \centering
  \includegraphics[width=0.9\linewidth]{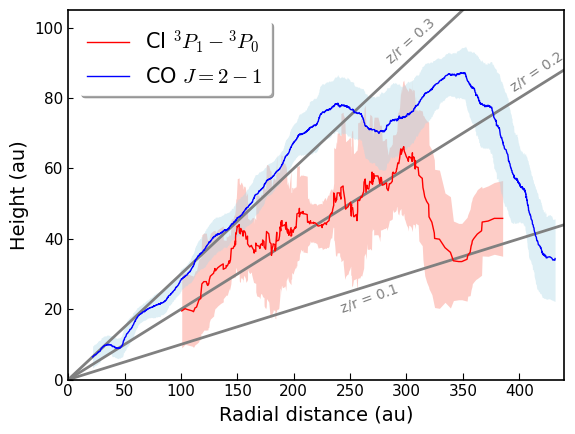}
\end{subfigure}%
\begin{subfigure}{.5\textwidth}
  \centering
  \includegraphics[width=.9\linewidth]{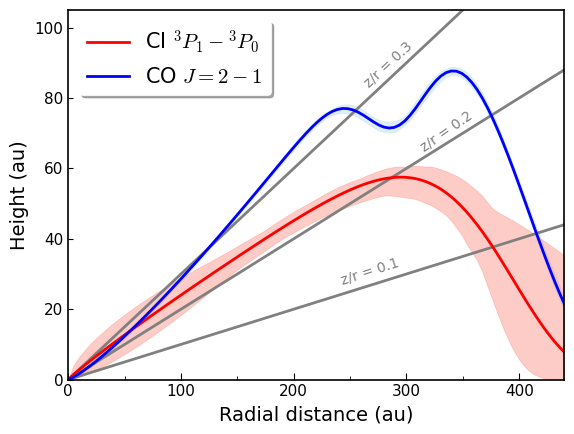}
\end{subfigure}
\caption{\textbf{Left:} average surface emission height plotted against deprojected radial distance from the star of both CO and [CI] obtained with \textbf{disksurf}. Shaded region shows $1\sigma$ scatter of the original obtained points shown in Fig. \ref{fig: residuals and data points}.  \textbf{Right:} best fitted model for [CI] and CO surface emission. For the first, a simple exponentially tapered powerlaw function (as described in eq. (\ref{eq: exponentially tapered})) is used, meanwhile for CO, a double exponentially tapered powerlaw function is considered to get a better modelling of its double peaked shape. Shaded region obtained by sampling posterior distribution of parameters. }
\label{fig: surface heigth}
\end{figure*}

\subsection{Vertical analysis}



As [CI] observations are spatially resolved, we attempt to extract properties of the vertical structure by using \textbf{disksurf} 
 \citep{disksurf} implementation of the method first presented in \cite{pintemethod}.  We use \textbf{disksurf} iterative approach that for each run updates a mask to better constrain the emission and better identify the peaks, in particular we used 5 and 10 iterations for obtaining CO and [CI] surface emission respectively.  Because of the low resolution of the [CI] observations the bottom side of the disk is not resolved, therefore this method in only applied to the upper surface  of the disk. The same is done for carbon monoxide, in a similar fashion as in \cite{Tere}.  

 Figure \ref{fig: surface heigth} left shows the obtained surface height for both [CI] and CO. [CI] has a ratio $z/r \sim 0.2$ up to 300 au, for greater radii it shows a drastic fall. CO is located in the interval 
 $z/r$ from 0.2 to 0.3, showing carbon monoxide tracing, on average, a higher layer of the surface, which seems to contradict our assumption that [CI] should be found close to the PDR at the surface of the disk. CO surface height shows a double peaked shape, also recovered by \cite{Tere}. The local minimum in the height of the CO emission is located at the same radial position with the maximum in the [CI] emission. 
 
 Given this method is depends on the spatial resolution (see section \ref{sec:angular res}), we repeated the vertical analysis in images produced using Brights weighting instead of natural. We were not able to recover the emission surface for values of robust parameter lower than 1 due to the low S/N. All profiles obtained with robust greater than 1 agreed with Fig. \ref{fig: surface heigth}.

\section{Discussions}
\label{sec:disucussions}
Figure \ref{fig:assymetry} shows an excess flux coming from the velocity channels from 5.2 to 7.3 km/s, also observed in the emission  enclosed by the 15$\sigma$ contours in the zeroth moment map shown in Fig. \ref{fig:zeroth}, that extends up to approximately 100 au from the central star.

In \cite{https://doi.org/10.48550/arxiv.2211.16531} a 2 km/s red-shifted component was first detected in [CI] observations at a  location of 0.5" to the SW of the central star and was proposed to be due to inflow of gas into the midplane. We were not able to measure the same asymmetry in the line even though an excess towards the high velocity channels was shown in Fig. \ref{fig:assymetry}, but we note that in \cite{https://doi.org/10.48550/arxiv.2211.16531} they also included data from an additional, shorter integration observations as part of project 2015.1.01137.S (PI: T. Tsukagoshi).  

As modelled in \cite{kama2016observations}, atomic carbon is expected to be located in the upper layers of the disk, tracing the PDR. We expect carbon monoxide and atomic carbon to be collocated radially, though not necessarily vertically, where atomic carbon may be tracing higher layers in the PDR.

Results shown in Fig. \ref{fig: PV diagrams} suggest a radial co-location of [CI] and CO from 40 au to 500 au. CO is  extended up to 600 au. The difference of 100 au may be explained by a lack of sensitivity to the weak [CI] emission in the outer disk in comparison to the deep CO observations. Notice equation (\ref{radial vel}) assumes a flat circular disk in Keplerian rotation, which is a good approximation at low ($r< 50$ au )and high radii ($r >$ 400 au) as shown in Fig. \ref{fig: surface heigth}, making the method valid for the purposes of constraining gas radial extent.

Regarding the surfaces shown in Fig. \ref{fig: surface heigth} left, we fitted both surface emissions by using exponentially tapered power law function defined by 

\begin{equation}
    z(r) = z_0\left(\frac{r}{r_0}\right)^{\psi} \times \exp\left(-\left[\frac{r}{r_{t}}\right]^{q_t}\right),
\label{eq: exponentially tapered}
\end{equation} 

where $r_0$ is just a constant fixed for units to match. For CO, given its double peaked shape, we followed a similar approach as in \cite{law2021molecules}, but we considered a baseline model given  by eq. \eqref{eq: exponentially tapered} and employed a Gaussian function for modelling the local minimum at $\sim 270$ au, therefore, for CO the parameter space consists on a total of 7 parameters, 4 from the baseline model and 3 more for the Gaussian feature to model the local minimum.  By using \textbf{emcee} package \citep{2013PASP..125..306F}, Markov chain Monte Carlo sampling is performed to explore the distribution of parameters whose posterior distributions are shown in 
Figs. \ref{fig:MCMC CO} and \ref{fig:MCMC CI}. 

Figure \ref{fig: surface heigth} right shows the best fitted models. In Fig. \ref{fig: residuals and data points} we show results of this fit, relative to the data points, and residuals of the model. The residuals range for [CI] goes from -40 to 40 au, which is within the ranges of the synthesized beam major axis.  Both the modelling and the data shown in Fig. \ref{fig: surface heigth} suggest atomic carbon to be, on average, located at lower altitudes compared to CO optically thick emission line  with ratio $z/r$ similar as other optically thin tracers like CN or HCN (see \cite{Tere} Fig. 2), with the first also matching [CI] maximum height at the same radius.  This is consistent with CN being also an external UV exposed regions tracer  (see \cite{bergner2021molecules} for a recent work about it). Notice HD 163296 is a Hearbig Ae star, system for which its disk [CI] line emission was predicted to be optically thin  (see \cite{kama2016observations} for general analysis in Hearbig Ae/Be systems), which may be able to explain the difference. 

Now, we set to explore this unexpected result by comparing it with physical-chemical models and proposing 3 different  effects that can contribute to the apparent non-vertical co-location of CO and
[CI] emissions.\begin{figure*}[t]
\centering
\begin{subfigure}{.5\textwidth}
  \centering
  \includegraphics[width=0.9\linewidth]{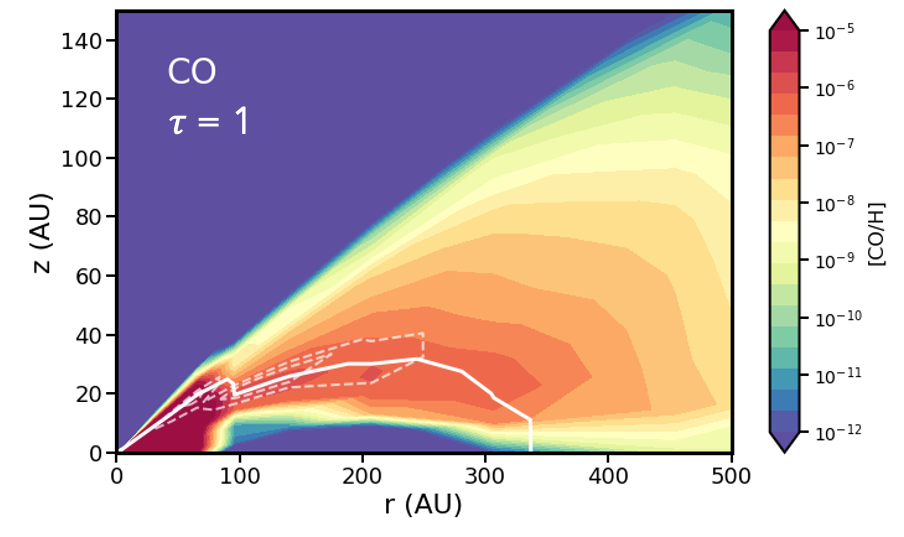}
\end{subfigure}%
\begin{subfigure}{.5\textwidth}
  \centering
  \includegraphics[width=.9\linewidth]{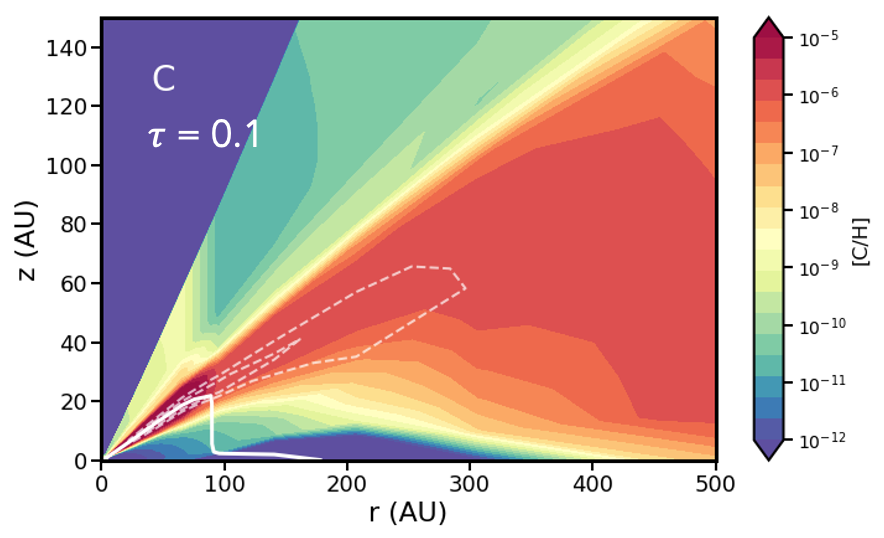}
\end{subfigure}
\caption{Abundance maps  for CO an [CI] in HD 163296 obtained with \textbf{DALI} modelling.  White dashed lines enclose areas with 75\% and 50\% of the emission expected for CO $J=2-1$ and [CI]$={^3}{P_1} - {^3}{P_0}$. White solid line shows line emission $\tau = 1$ and $\tau=0.1$ surfaces for CO and [CI] line emission respectively.  }
\label{fig: abundance maps}
\end{figure*}

\subsection{DALI modelling of [CI] and CO in HD 163296}

To compare our current understanding about the formation of [CI] and its relation to CO in HD 163296, we make use of the \textbf{DALI} physical-chemical modelling code \citep{bruderer2012warm, 2013A&A...559A..46B} to obtain CO and [CI] abundance vertical profiles as well as their expected emission layers. Starting with a parameterized gas and dust density distribution and an input stellar spectrum, the code utilizes Monte Carlo radiative transfer to calculate the UV radiation field and dust temperature. This establishes an initial guess for the gas temperature, which serves as the starting point for an iterative process to solve gas-grain chemistry (time-dependently or in equilibrium). Spectral image cubes, line profiles, and disk-integrated line fluxes are then be obtained using the ray-tracing module. In appendix \ref{appendix: dali vs disksurf}, we show how the image cubes emission surface , extracted with \textbf{disksurf}, compares to the \textbf{DALI} emission contours to test how well this method is able the extract the vertical structure of the gas.  
The model is fit following the procedure outlined in \citet{kama2016volatile} and \citet{keyte_2023}, see those works for a full description of the fitting process. In short, the physical structure and total dust mass is first determined by fitting the continuum spectral energy distribution using a large grid of models. The parameters $R_\text{gap}$, $\psi$, $h_\text{c}$, $\delta_\text{dust}$, and $\Delta_\text{g/d}$ are varied, while $\Sigma_\text{c}$ is kept fixed at an arbitrary value such that changes in $\Delta_\text{g/d}$ are equivalent to changes only in the total dust mass. Having determined $M_\text{dust}$, we adopt a gas-to-dust ratio of $\Delta_\text{g/d}=100$, and fix $\Sigma_\text{c}$ such that the best-fit dust mass is maintained. The total elemental carbon and oxygen abundances are then constrained by fitting the spatially unresolved CO rotational ladder and resolved CO 3-2 observations. Final molecular abundances are obtained using a chemical network based on subset of the UMIST 06 network \citep{woodall_umist}, which consists of 118 species and 1675 individual reactions. We run the chemistry time-dependently to a chemical age of 1 Myr, which is typical of such systems.  A summary of the derived parameters is reported in Table \ref{table:modelparameters}.

From the image cubes we were not able to reproduce the asymmetry in the emission that enhanced the near side of the disk, therefore, we can exclude radiative transfer effects creating the profile showed in Fig. \ref{fig:assymetry}. Notice this phenomena can't be explained by geometry as we have projected the emission with the same inclination and PA as the data.

Figure \ref{fig: abundance maps} shows the results for the modelling in HD 163296. There CO emission seems to follow the curve of $\tau=1$ as expected. For [CI] we overlaid the curve of $\tau =0.1$ as higher optical depths are obtained only at the midplane and therefore its line emission is optically thin as previous modelling of Herbig Ae/Be systems suggested \citep{kama2016observations}.

\begin{figure}[h]
    \centering
    \includegraphics[scale=0.6]{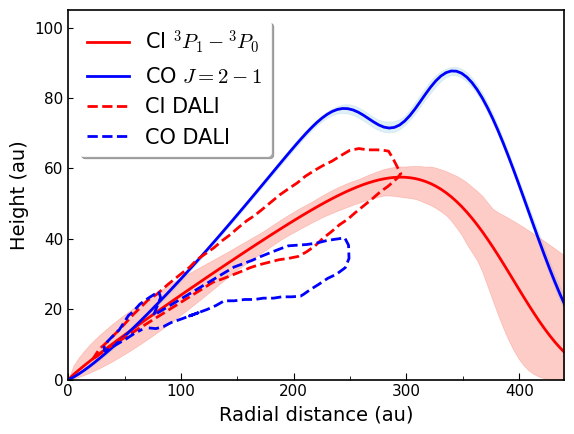}
    \caption{Same as Fig. \ref{fig: surface heigth} right but with the addition of 75\% line emission contours from HD 163296 DALI models.  } 
    \label{fig: DALI comparison}
\end{figure}

 Figure \ref{fig: DALI comparison} shows the comparison of the 75\%  emission contour with our derived surface emission for CO and [CI]. The DALI model  agrees with derived [CI] surface emission  but for CO it strongly disagrees as the predicted emission surface is located at much lower heights from radii higher than 100 au.

 We would like to stress that the \textbf{DALI} model reflects our current understanding about the PDR in protoplanetary disks as it predicts [CI] being in higher altitudes compared to CO where we expected UV radiation is the most efficient in protoplanetary disks.
 
 \begin{figure*}[t]
\centering
\begin{subfigure}{.5\textwidth}
  \centering
  \includegraphics[width=0.9\linewidth]{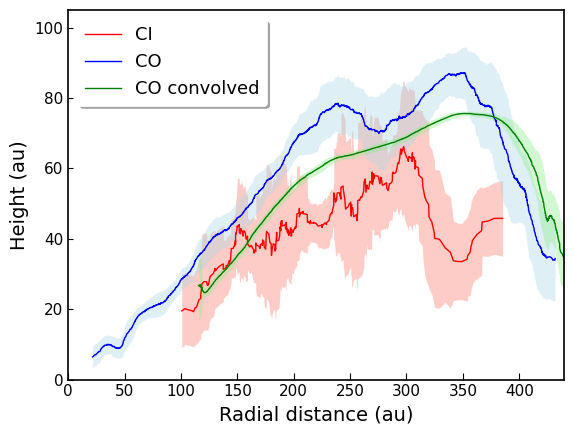}
\end{subfigure}%
\begin{subfigure}{.5\textwidth}
  \centering
  \includegraphics[width=.9\linewidth]{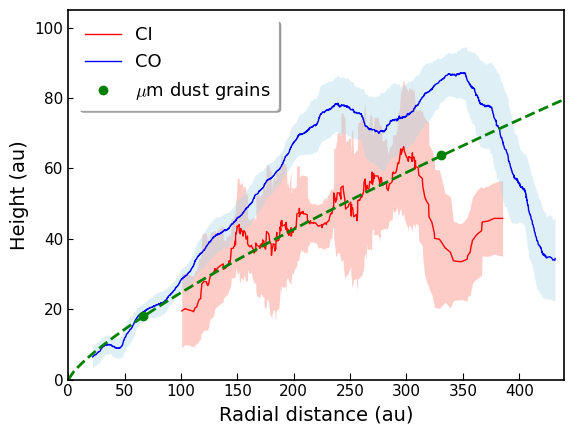}
\end{subfigure}
\caption{\textbf{Left:} same as Fig. \ref{fig: surface heigth} with the addition of the measured height in convolved CO data shown in green.  \textbf{Right:} same as Fig. \ref{fig: surface heigth} with the addition of a green dashed line and points showing $\mu m$ sized dust grains height power law fitting and data points from \cite{rich2021investigating} respectively.}
\label{fig:  surface heigth appendix}
\end{figure*}

\subsection{Angular resolution effects}
\label{sec:angular res}
The method employed in the present work for deriving vertical height depends on the correct location of local maxima from each channel. When the angular resolution is not enough to resolver both the upper and lower part of the disk, mixing of both emissions would be the observed. This effect could shift the local maximum towards the major axis of the disk which will result in a lower height measurement (see \cite{rosenfeld2013spatially} Fig. 4 for visualization of this effect). 

To test whether angular resolution can explain the observed discrepancy, the CO data is convolved with a 2D elliptical Gaussian with the same PA, minor and major axis as the synthesized beam of the [CI] observations in an attempt of mimicking the low angular resolution of the latter. Then, the vertical analysis is repeated and results are shown in Fig. \ref{fig: surface heigth appendix} left. 

From the convolved data we were not able to detect any point at radii minor than 100 au, which is also seen in the [CI] profile, indicating that it is not possible to resolve inner structure as expected from a low angular resolution case. The vertical profile of CO decreases on average a 10\% to 15\%  diminishing the difference between CO and [CI]. We employed a Welch's t-test to examine the null hypothesis that the convolved CO and [CI] data have the same average height within the 100 to 300 au range.
The analysis yielded a p-value lower than $10^{-10}$ and therefore rejecting the hypothesis with a certainty higher than $6\sigma$, suggesting that resolution effects alone may not be sufficient to explain this difference.

\subsection{$\mu$m dust grains vertical non co-location with CO}

In gas and dust rich environments such as protoplanetary disks, the shielding from high energy photons is usually provided by $\mu$m dust grains \citep{henning2013chemistry}. If dust grains and CO are vertically co-located, we expect the PDR to be a thin structure tracing the surface of the disk. In this case we also expect [CI] to trace the same surface as the optically thick CO line. This lead us to hypothesize that the discrepancy observed may be due to $\mu$m dust grains being vertically non co-located with gaseous CO. This non co-location would produce a wedge like PDR region and the optically thin [CI] line would trace the transition from the shielded to non-shielded part of the disk. Notice that in this UV-exposed region we still need a high  abundance of CO such that the $J=2-1$ line remains optically thick or a high enough accumulation of [CI] such that CO gets shielded by the latter.

A recent work from \cite{rich2021investigating} tried to constrain the vertical profile small dust grains in 3 different protoplanetary disks, HD 163296, IM Lup and CU Cha, finding that for the first two, small grains were located below the CO gas height, which may be produced by a large gas-to-dust ratio that lowers the scattering surface heights.  For comparison purposes, in Fig. \ref{fig: surface heigth appendix} right, we added to our height profiles the power law fitting of dust grains as well as the data points from the same study. Even though the vertical profile for the dust grains in HD 163296 is constrained by only two data points, it matches the [CI] profile providing direct observational evidence in favor of our hypothesized PDR wedge like structure.

Furthermore, under the assumption the PDR vertical structure depends primarily on $\mu$m dust grains position, systems like IM Lup and CU Cha offer a unique opportunity to test its diversity by using future high angular resolution [CI] observations. For CU Cha, as both CO and dust grains profiles show a good agreements, we expect a the [CI] and CO to also agree in their vertical profile.  From \cite{rich2021investigating} work, IM Lup vertical profiles show an even more drastic difference between CO and small dust grains compared to HD 163296, therefore  if the main mechanism for UV shielding in IM Lup came from $\mu m$ dust grains, we would have expected this system to show a similar vertical location of [CI] emission relative to CO as the one observed in this work for HD 163296, but in \cite{Imlup_recent} it was observed exactly the opposite phenomena in which [CI] appears to trace upper layers with $z/r \geq 0.5 $, higher than for CO with $z/r \approx 0.3-0.4$. This difference may come from the gas abundance and gas-to-dust ratio of both systems as IM Lup was found to be a massive disk with mass of $\sim 0.1 M_{\odot}$  \citep{2008A&A...489..633P}, in contrast, HD 163296 has gaseous content with mass of $\sim 0.06 M_{\odot}$  as reported in Table \ref{table:modelparameters}. 

\subsection{Mixing of upper layers}

Finally, [CI] may be located at a lower altitude than CO due to an efficient mechanism of mixing in the upper layers of the disk, making [CI] on average, to be located at a lower height and not exclusively at the surface of the disk. One way in which this may happen is through meridional flows which are expected to arise in the early stages of planet formation \citep{morbidelli2014meridional}. This would be dependent on the mixing time and the destruction timescale of CO.

In \cite{teague2019meridional}, high precision velocity signatures, resembling meridional flows, were detected in HD 163296. These flows match the position of previously detected continuum ring gaps and a kinematic feature detected in CO observations, postulated to be due to the presence of a 2 Jupiter-mass planet \citep{pinte2018kinematic}. As this phenomena seems to be localized at certain locations within the disk, its effects get diluted by the method employed in this work which assumes an azimuthal symmetry of the disk. Therefore, we expect this effect to contribute the less compared to the other two we explored before . 
\section{Conclusions}
\label{sec:summary}

By using spatially resolved, archival ALMA observations of the [CI]$={^3}{P_1} - {^3}{P_0}$ emission line towards HD 163296, we have made the first analysis of both the radial and vertical location of atomic carbon in a class II protoplanetary disk and performed the first direct comparison of this tracer with carbon monoxide by using observations of $^{12}$CO $J=2-1$ optically thick emission line. We found [CI] shows an excess of flux towards the near side of the disk with a radial extension of 100 au which may be related to planets candidates. CO and [CI] are collocated radially up to a radius of 500 au, with first being more radially extended, this difference may be explained by the low sensitivity achieved in [CI] observations. 

With radial average profiles of both dust continuum and the zeroth moment map of [CI], we were able to compare the radial extent of the [CI] emission with dust continuum finding the first is 2.6 times more radially extended than dust emission, in good agreement with the fact HD 163296 is a gas rich protoplanetary disk.

Regarding the vertical structure of the selected tracers, CO shows a double peaked surface emission, whose ratio $z/r$ lies in the interval $0.3 - 0.2$ before decreasing towards the outer disk at a radius of 350 au. For [CI], despite only 13.1 minutes of integration time were used to obtain the data, a  vertical structure with a ratio  $z/r \sim 0.2$ is measured up to 300 au,  comparable to tracers like CN or HCN, but located, on average, at lower altitude than CO. 

Brand new \textbf{DALI} models were obtained for HD 163296 and its CO and [CI] abundances and emission. The model shows CO emission follows closely the $\tau = 1$ curve as expected in contrast to [CI], whose emission is optically thin. The derived surface emission
of [CI] agrees with the model but CO  emission surface is heavily underestimated by it. 
Three different effects were explored in order to explain the unexpected low heights of [CI] emission compared to CO. First, low angular resolution of [CI] data may have decreased the measured profile, but it is not able to completely explain the discrepancy as the CO convolved data profile still showed a significant difference. Second, a wedge-like PDR structure instead of a thin surface towards the outer part of the disk may be able to explain the difference, which is supported by observational evidence of $\mu$m dust grains being non co-located vertically with CO as the latter account for most of the shielding in dust-gas rich environments. Finally, mixing of the upper layers may also contribute which could be done by the meridional flows observed in HD 163296.

Future observations of [CI] towards protoplanetary disks will enable exploring the PDR in detail, where targets like IM Lup and CU Cha offer a good starting points. New theoretical modelling for the PDR of protoplanetary disks will also be needed ideally including the effects of dust distribution, mixing and shielding.

\begin{acknowledgements}
      Francisco Urbina acknowledges Subdirection of Human Capital ANID for providing fundings for his MSc program (national MSc 2023/22231861).  Mihkel Kama gratefully acknowledges funding from the European Union's Horizon Europe research and innovation programme under grant agreement No. 101079231 (EXOHOST), and from UK Research and Innovation (UKRI) under the UK government’s Horizon Europe funding guarantee (grant number 10051045). Luke Keyte acknowledges funding via a Science and Technology Facilities Council (STFC) studentship.
      We thank the referee as the comments and questions during the revision process made the article significantly better. 
\end{acknowledgements}

\bibliographystyle{aa} 
\bibliography{article_bib} 
\begin{appendix}
\section{Channel maps [CI] emission}

\begin{minipage}{1.0\textwidth}
  \strut\newline
  \centering
 \includegraphics[scale=0.6]{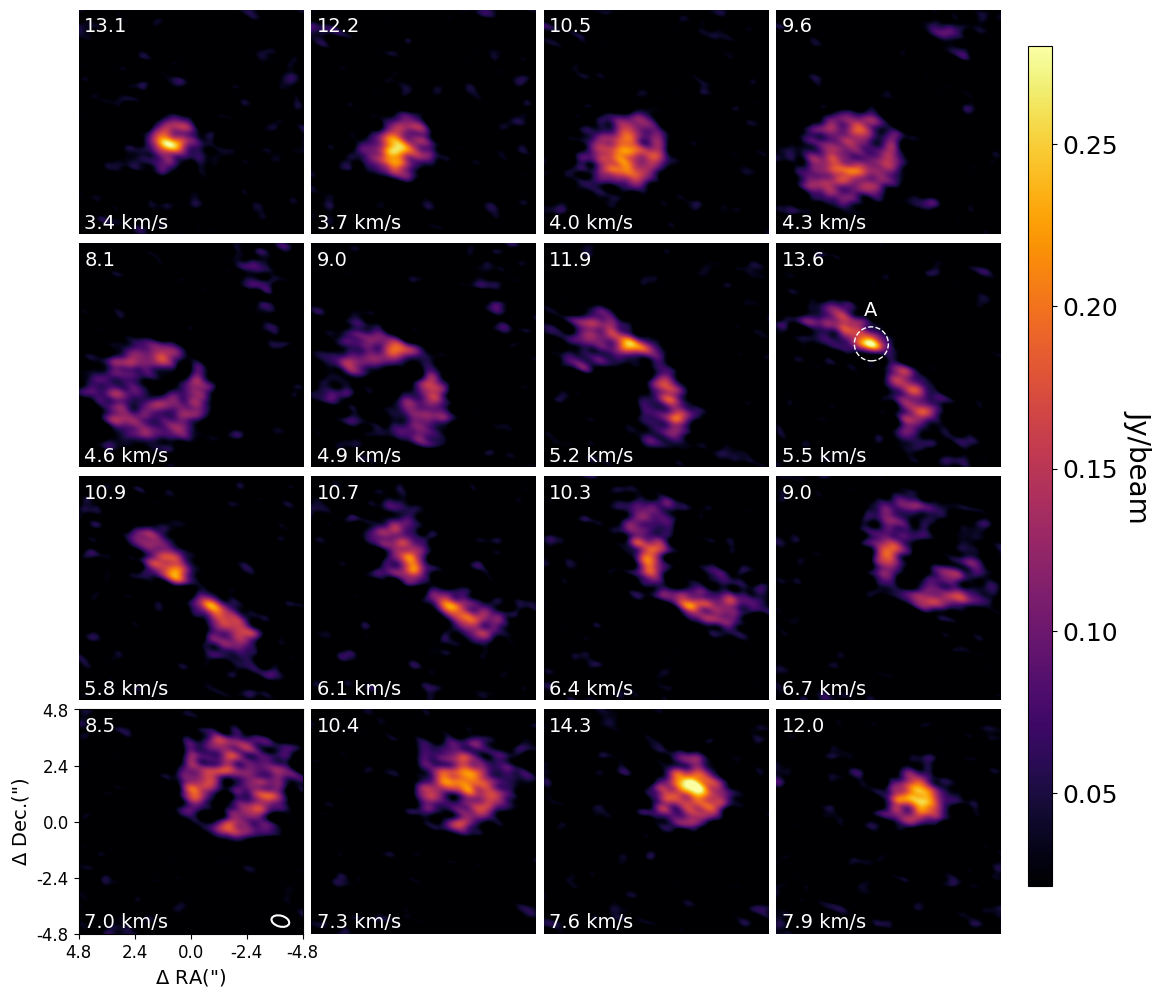}
    \captionof{figure}{[CI]$={^3}P_1 - {^3}P_0$ line emission channel maps observed in HD163296. Each panel has dimensions of $9.64 " \times 9.64 "$, in the bottom left is labeled by its velocity relative to the rest frequency of the [CI] line. In the top left the peak SNR is shown. Panel of 7.0 km/s shows the synthesized beam FWHM  with a white ellipse. The enclosed region at the panel of 5.5 km/s, labelled by the letter A, highlights an asymmetry feature in the emission.}
    \label{fig:channels}
\end{minipage}
\clearpage

\section{exponential taper modeling residuals and posterior distribution of parameters}

\label{appendix: modeling}

\begin{minipage}{1.0\textwidth}
  \strut\newline
  \centering
  \includegraphics[scale=0.45]{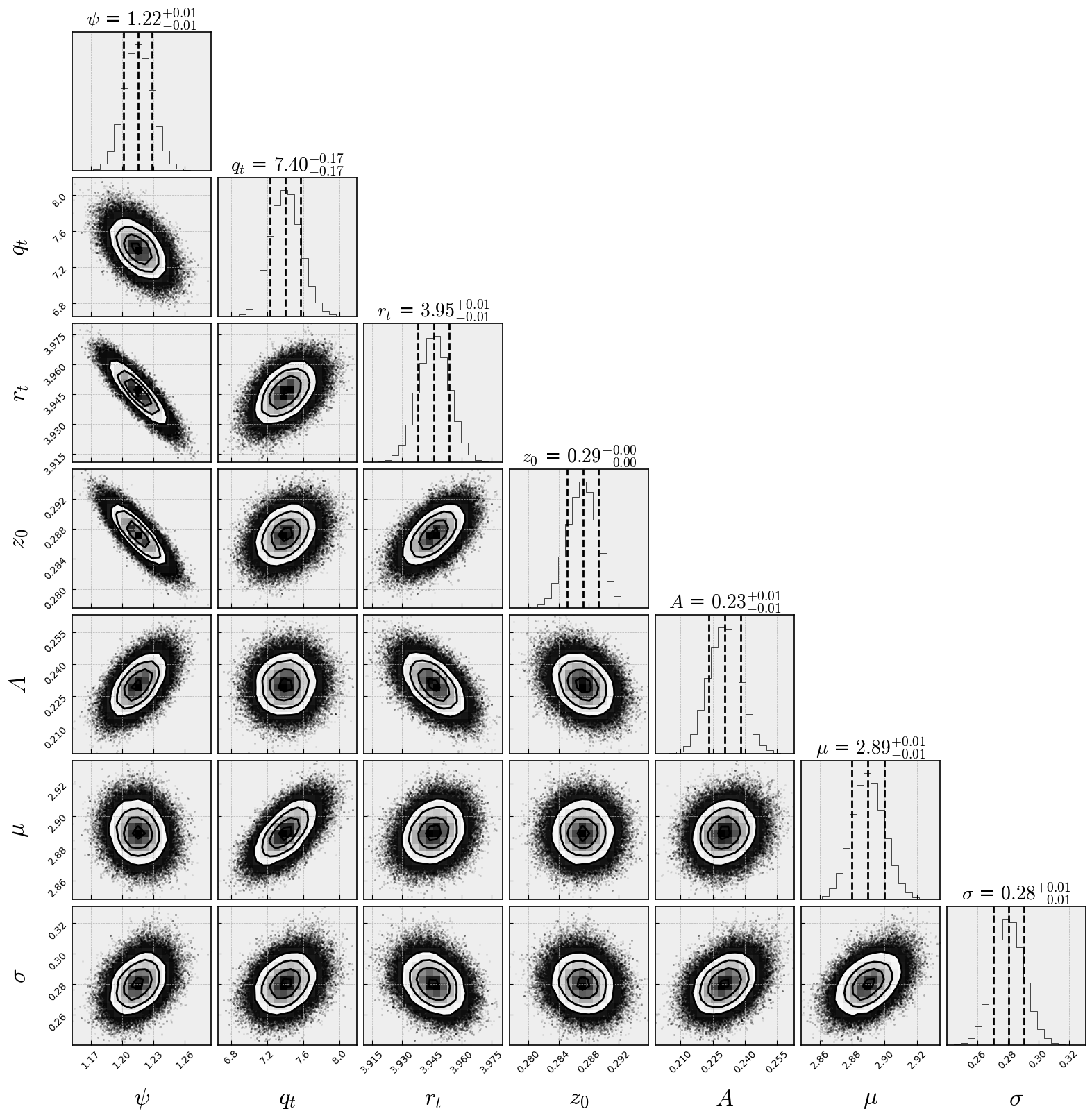}
    \captionof{figure}{Posterior distribution for parameters $\psi, q_{t}, r_{t}$ and $z_0$  described in eq. (\ref{eq: exponentially tapered}) and  $A, \mu$ and $\sigma$ from the Gaussian depression for $^{12}$CO $J=2-1$ data. Marginalized distributions are shown in the diagonal and their vertical dashed lines represent the 16th, 50th and 84th percentiles.}
    \label{fig:MCMC CO}
\end{minipage}

\begin{figure*}[h]
    \centering
    \includegraphics[scale=0.7]{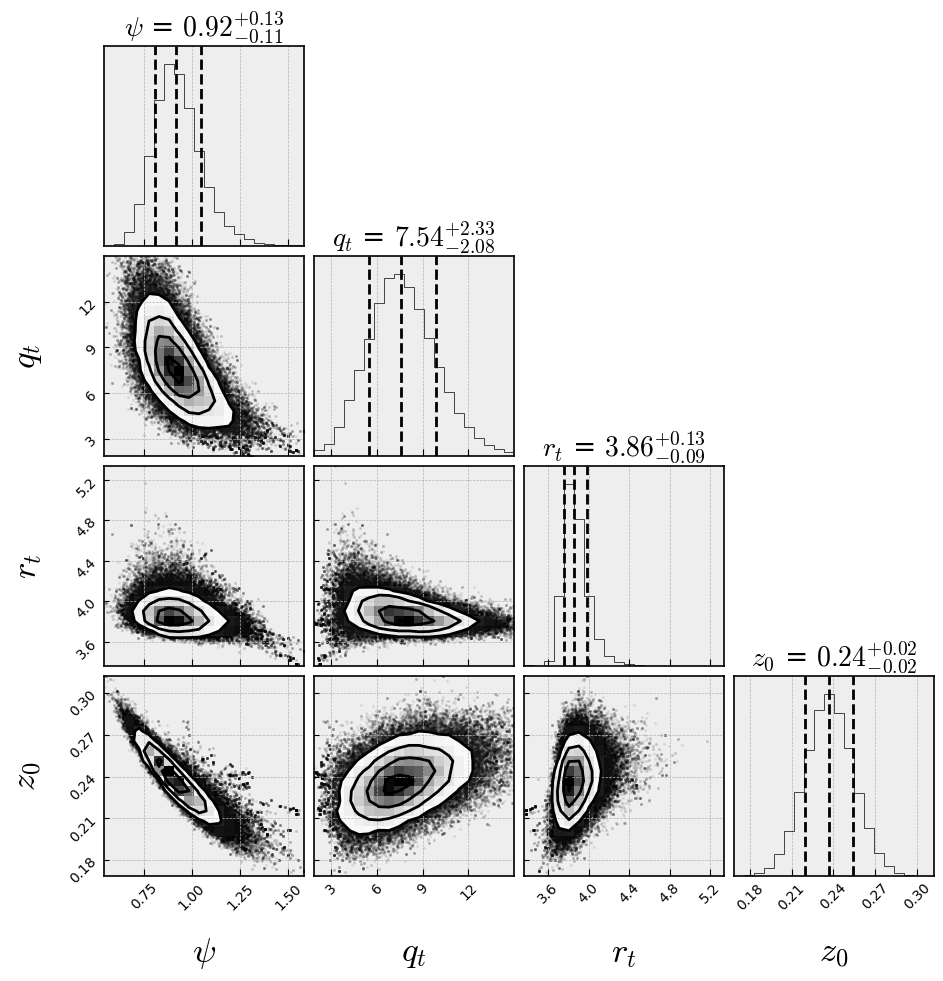}
    \caption{Posterior distribution for parameters $\psi, q_{t}, r_{t}$ and $z_0$ from eq. (\ref{eq: exponentially tapered}), for fitting surface height in [CI] data. Marginalized distributions are shown in the diagonal and their vertical dashed lines represent the 16th, 50th and 84th percentiles.}
    \label{fig:MCMC CI}
\end{figure*}

\begin{figure*}[h]
\centering
\begin{subfigure}{.5\textwidth}
  \centering
  \includegraphics[width=.85\linewidth]{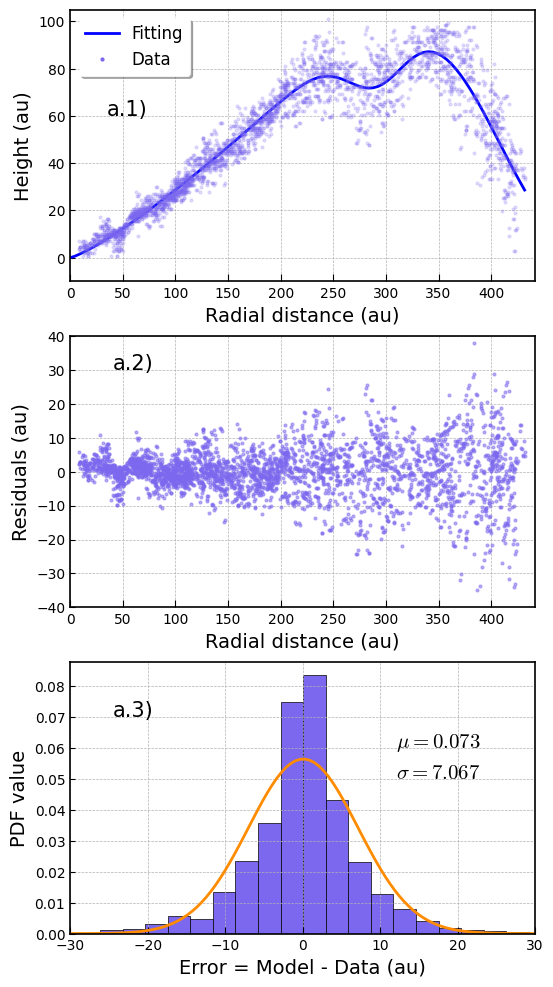}
  \caption{Modeling and residuals for CO vertical analysis.}
\end{subfigure}%
\begin{subfigure}{.5\textwidth}
  \centering
  \includegraphics[width=.85\linewidth]{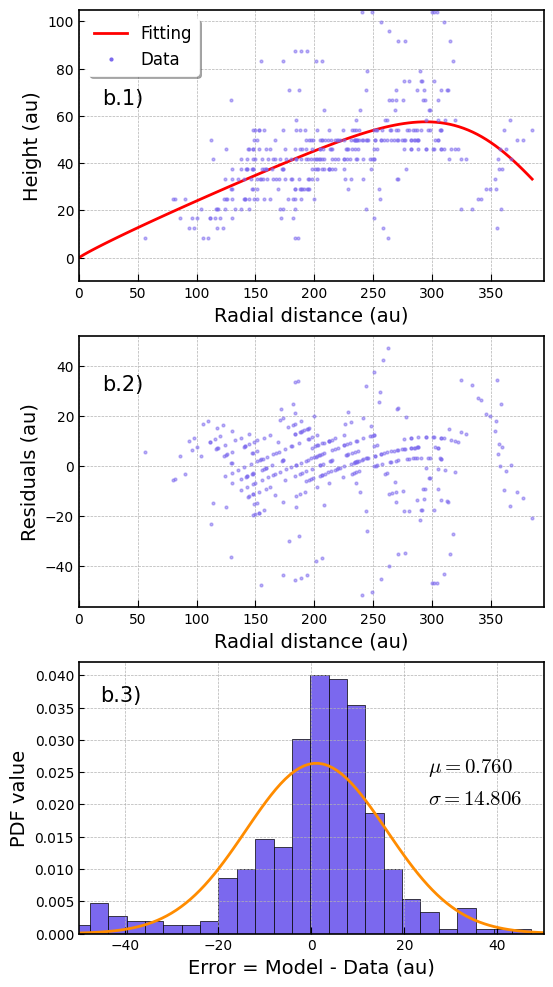}
  \caption{Modeling and residuals for [CI]
 vertical analysis.}
\end{subfigure}
\caption{(a.1, b.1): Purple data points show the surface height obtained with  \textbf{disksurf}, applied to the data presented in \ref{sec:observations}. Blue and red solid lines show the best fitted model for CO and [CI] respectively. (a.2, b.2): Residuals as a function of radial distance. (a.3, b.3): Residuals normalized histogram. Orange solid line shows a normal distribution with mean and standard deviation measured from residuals. }
\label{fig: residuals and data points}
\end{figure*}

\clearpage

\section{DALI model parameters}
\label{appendix:dali_model}
\begin{table}[!htbp]
\caption{Fiducial HD~HD169296 disk model parameters.}             
\label{table:modelparameters}      
\centering
\begin{tabular}{l l l}     %
\hline\hline       
                      
Parameter & Description & Value\\ 
\hline                    
   $R_\text{sub}$       & Sublimation radius                         & 0.41 au    \\
   $R_\text{gap}$       & Inner disk size                            & 3.66 au    \\
   $R_\text{cav}$       & Cavity radius                              & 3.67 au    \\
   $R_c$                & Critical radius for surface density        & 125 au     \\
   $R_\text{out}$       & Disk outer radius                          & 1000 au    \\
   $\delta_\text{gas}$  & Gas depletion factor inside cavity         & 1.0        \\
   $\delta_\text{dust}$ & Dust depletion factor inside cavity        & 0.01       \\
   $\gamma$             & Power law index of surface density profile & 0.9        \\
   $\chi$               & Dust settling parameter                    & 0.2        \\
   $f$                  & Large-to-small dust mixing parameter       & 0.9        \\
   $\Sigma_c$           & $\Sigma_\text{gas}$ at $R_c$               & 6.8 g cm$^{-2}$  \\
   $h_c$                & Scale height at $R_c$                      & 0.075      \\
   $\psi$               & Power law index of scale height            & 0.05       \\
   $\Delta_\text{gd}$   & Gas-to-dust mass ratio                     & 100        \\
   $L_X$                & Stellar X-ray luminosity                   & $3.98 \times 10^{29} \text{ erg s}^{-1}$    \\
   $T_X$                & X-ray plasma temperature                   & $5.7 \times 10^{5}$ K           \\
   $\zeta_\text{cr}$    & Cosmic ray ionization rate                 & $1.7 \times 10^{-17}$ s$^{-1}$  \\
   $M_\text{gas}$       & Disk gas mass                              & $6.7 \times 10^{-2}M_{\odot}$      \\
   $M_\text{dust}$      & Disk dust mass                             & $6.6 \times 10^{-4}M_{\odot}$     \\
   $t_\text{chem}$      & Timescale for time-dependent chemistry     & \text{1 Myr}                    \\
\hline                  
\end{tabular}
\end{table}

\section{Simulated image cubes surface emission}
\label{appendix: dali vs disksurf}
To test how well \textbf{disksurf} tool traces the surface emission of the gas, we repeated the vertical analysis in the simulated image cubes obtained with the DALI modelling.  The results, compared to DALI 75\% emission contours, are shown in Fig. \ref{fig: DALI vs simulated cubes} in the appendix. There we notice that both the [CI] profile is located at a higher altitude from radii $>100$ au, but the difference is not as drastic as the DALI contours suggest. This is an effect of the emission being optically thin. The CO emission surface follows the DALI contours, which corresponds approximately to the $\tau = 1$ surface for the line.

 \begin{figure}[h]
     \centering
     \includegraphics[scale=0.6]{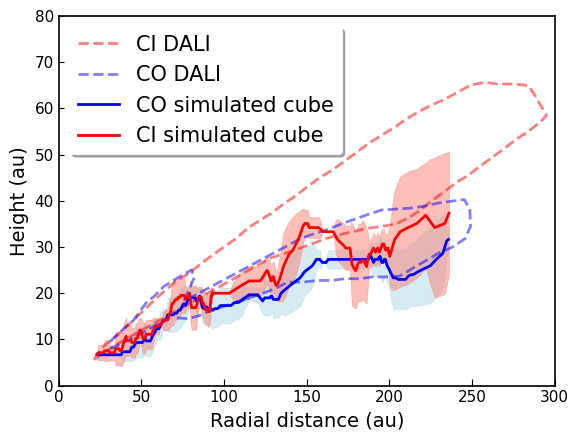}
     \caption{Similar as Fig. \ref{fig: DALI comparison} but with the extracted profiles of the simulated image cubes. }
     \label{fig: DALI vs simulated cubes}
 \end{figure}
\end{appendix}
\end{document}